\documentclass[a4paper,11pt]{article}
\pdfoutput=1 

\usepackage{jheppub} 

\usepackage[T1]{fontenc} 

\title{\boldmath Massive Celestial Fermions}

\author[a,1]{Sruthi A. Narayanan\note{Corresponding author.}}

\affiliation[a]{Center for the Fundamental Laws of Nature, Harvard University,\\Cambridge, MA 02138, USA}

\emailAdd{sruthi\_narayanan@g.harvard.edu}

\abstract{In an effort to further the study of amplitudes in the celestial CFT (CCFT), we construct conformal primary wavefunctions for massive fermions. Upon explicitly calculating the wavefunctions for Dirac fermions, we deduce the corresponding transformation of momentum space amplitudes to celestial amplitudes. The shadow wavefunctions are shown to have opposite spin and conformal dimension $2-\Delta$. The Dirac conformal primary wavefunctions are delta function normalizable with respect to the Dirac inner product provided they lie on the principal series with conformal dimension $\Delta = 1+i\lambda$ for $\lambda\in\mathbb{R}$. It is shown that there are two choices of a complete basis: single spin $J=\frac{1}{2}$ or $J=-\frac{1}{2}$ and $\lambda\in\mathbb{R}$ or multiple spin $J=\pm\frac{1}{2}$ and $\lambda\in\mathbb{R}_{+\cup 0}$. The massless limit of the Dirac conformal primary wavefunctions is shown to agree with previous literature. The momentum generators on the celestial sphere are derived and, along with the Lorentz generators, form a representation of the Poincar\'e algebra. 
Finally, we show that the massive spin-$1$ conformal primary wavefunctions can be constructed from the Dirac conformal primary wavefunctions using the standard Clebsch-Gordan coefficients. We use this procedure to write the massive spin-$\frac{3}{2}$, Rarita-Schwinger, conformal primary wavefunctions. This provides a prescription for constructing all massive fermionic and bosonic conformal primary wavefunctions starting from spin-$\frac{1}{2}$.}

\begin{document} 
\maketitle
\flushbottom

\section{Introduction}
Recently there has been considerable focus on the conformal primary basis. The transformation from a plane wave basis to the conformal primary basis provides a mapping from 4-dimensional (4D) momentum space scattering amplitudes to 2-dimensional (2D) "celestial amplitudes". These celestial amplitudes transform as conformal correlators under $\mbox{SL}(2,\mathbb{C})$. The conformal primary basis was introduced
in~\cite{Pasterski:2016qvg,Pasterski:2017kqt} for amplitudes containing massless and massive scalars, photon and graviton fields. It has since been extended to all massless fields in~\cite{Pasterski:2017ylz,Lam:2017ofc,Fotopoulos:2020bqj}. This construction was then generalized to massive bosons in~\cite{Law:2020tsg}. It has been further used to study the analytic structure of amplitudes as well as conformally soft theorems in~\cite{Donnay:2018neh,Pate:2019lpp,Puhm:2019zbl, Law:2019glh, Pate:2019mfs,Banerjee:2020zlg,Law:2020xcf, Casali:2020vuy, Albayrak:2020saa}. 

The correspondence between a 4D bulk theory and a 2D boundary conformal field theory has its roots in earlier work. In~\cite{deBoer:2003vf}, de Boer and Solodukhin introduced a 3-dimensional (3D) hyperbolic slicing of 4D Minkowski space where the common boundary of the slices was a 2-sphere, the celestial sphere, $\mathcal{CS}$. This led to an important holographic relation between bulk Lorentz symmetries and conformal symmetries on $\mathcal{CS}$ as demonstrated in~\cite{Kapec:2014opa,Kapec:2017gsg}. The authors of~\cite{Cheung:2016iub} were then able to recast 4D scattering amplitudes as 2D conformal correlators which have come to be referred to as celestial amplitudes.

To widen the set of amplitudes that we can study using the conformal primary basis, it is a natural next step to understand the transformation of fermionic fields. Whereas it is easier, in principal, to generalize transformations for massless particles because, in a particular gauge, they are Mellin transforms of plane wave solutions, massive fields are trickier and have a different functional form for each spin. The goal of this paper is to explicitly construct the spin-$\frac{1}{2}$ (Dirac) conformal primary wavefunctions and to show that they can be used to construct the spin-1 primaries. This results in a prescription to construct massive conformal primary wavefunctions for arbitrary integer and half integer spin via the standard Clebsch-Gordan coefficients.

This paper is organized as follows. In section~\ref{sec:prelims} we detail the conventions used throughout this paper. In particular, we review pertinent recent results for conformal primary wavefunctions of massive bosons and the Dirac equation in Minkowski space with $(-,+,+,+)$ signature. In section~\ref{sec:wavefunctions} we explicitly construct the Dirac conformal primary wavefunctions using constraints and discuss how to transform an amplitude containing fermions from momentum space to a celestial amplitude.  In section~\ref{sec:analytic} we write the explicit functional form of the wavefunctions in terms of the previously found scalar conformal primaries. We compute the shadow field and show that it shifts the conformal dimension from $\Delta$ to $2-\Delta$ as well as flipping the spin. We compute the Dirac inner product of two conformal primary wavefunctions to demonstrate that they are delta function normalizable as long as we consider the principal continuous series $\Delta=1+i\lambda$ for $\lambda\in\mathbb{R}$. We note that there are two choices for a complete basis: $J=\frac{1}{2}$ or $J=-\frac{1}{2}$ and $\lambda\in\mathbb{R}$ or $J=\pm\frac{1}{2}$ and $\lambda\in\mathbb{R}_{+\cup 0}$. Finally we compute the massless limit and show it is in agreement with the spinor solutions in~\cite{Fotopoulos:2020bqj}. In section~\ref{sec:momentum} we derive the spin$-\frac{1}{2}$ momentum generators in the celestial basis and show that they are not diagonal and that they form a representation of the Poincar\'e algebra along with the Lorentz generators. In section~\ref{sec:spin1} we arrive at a prescription for writing down arbitrary massive half-integer and integer spin conformal primary wavefunctions by first relating the spin-$\frac{1}{2}$ to the spin-$1$ conformal primary wavefunctions and then constructing the Rarita-Schwinger fields, massive spin-$\frac{3}{2}$, using Clebsch Gordan coefficients. In appendix~\ref{app:diffeq} we write the details of solving the differential equations to get the wavefunctions. In appendix~\ref{app:bulkint} we compute a weighted bulk integral of scalar propagators which is useful for the Dirac inner product. In appendix~\ref{app:boundary} we compute a boundary integral of scalar propagators which is useful for the completeness relation.

After this work was completed, independent and overlapping results were posted in~\cite{Muck:2020wtx}.

\section{Preliminaries}\label{sec:prelims}
We start with a hyperbolic slicing of 4D Minkowski spacetime with signature $(-,+,+,+)$ where the metric on each 3D hyperbolic slice, $\mathcal{H}_3$, is given by 
\begin{equation}
ds^2 = \frac{dy^2+dzd\bar{z}}{y^2}.
\end{equation}
The coordinate $y$ varies as we move along $\mathcal{H}_3$. When $y\rightarrow 0$ we approach $\mathcal{CS}$, parametrized by complex coordinates $z,\bar{z}$. The momentum of a massive particle is parametrized as 
\begin{equation}
p_k^\mu = m_k\left(\frac{1+y_k^2+z_k\bar{z}_k}{2y_k},\frac{\bar{z}_k+z_k}{2y_k}, \frac{i(\bar{z}_k-z_k)}{2y_k}, \frac{1-y_k^2-z_k\bar{z}_k}{2y_k}\right)\equiv m_k\hat{p}_k^\mu
\end{equation}
where $p_k^2 = -m_k^2$. Massless (null) momenta are parametrized as 
\begin{equation}
q^\mu = \omega (1+w\bar{w},\bar{w}+w,i(\bar{w}-w),1-w\bar{w})\equiv \omega\hat{q}^\mu.
\end{equation}
Each hyperbolic slice has an SL$(2,\mathbb{C})$ isometry given by the coordinate transformations 
\begin{equation}
\label{coordtrans}
z\rightarrow z' = \frac{(az+b)(\bar{c}\bar{z}+\bar{d})+a\bar{c}y^2}{(cz+d)(\bar{c}\bar{z}+\bar{d})+c\bar{c}y^2}, \ \ \ \ y\rightarrow y' = \frac{y}{(cz+d)(\bar{c}\bar{z}+\bar{d})+c\bar{c}y^2}
\end{equation}
where $ad-bc=\bar{a}\bar{d}-\bar{b}\bar{c}=1$. The above SL$(2,\mathbb{C})$ transformation induces M\"obius transformations of the complex coordinates $w,\bar{w}$ on $\mathcal{CS}$
\begin{equation}
w\rightarrow w'=\frac{aw+b}{cw+d}, \ \ \bar{w}\rightarrow \bar{w}'=\frac{\bar{a}\bar{w}+\bar{b}}{\bar{c}\bar{w}+\bar{d}}.
\end{equation}
The appropriate Lorentz transformation matrix is given explicitly~\cite{Oblak:2015qia} by 
\begin{equation}
\begin{small}
\Lambda^\mu_\nu = \frac{1}{2}\begin{pmatrix} a\bar{a}+b\bar{b}+c\bar{c}+d\bar{d} & b\bar{a}+a\bar{b}+d\bar{c}+c\bar{d} & i(-b\bar{a}+a\bar{b}-d\bar{c}+c\bar{d}) & -a\bar{a}+b\bar{b}-c\bar{c}+d\bar{d}\cr c\bar{a}+a\bar{c}+d\bar{b}+b\bar{d} & d\bar{a}+a\bar{d}+c\bar{b}+b\bar{c} & i(-d\bar{a}+a\bar{d}+c\bar{b}-b\bar{c}) & -c\bar{a}-a\bar{c}+d\bar{b}+b\bar{d}\cr i(c\bar{a}-a\bar{c}+d\bar{b}-b\bar{d}) & i(d\bar{a}-a\bar{d}+c\bar{b}-b\bar{c}) & d\bar{a}+a\bar{d}-c\bar{b}-b\bar{c} & i(-c\bar{a}+a\bar{c}+d\bar{b}-b\bar{d})\cr -a\bar{a}-b\bar{b}+c\bar{c}+d\bar{d} & -b\bar{a}-a\bar{b}+d\bar{c}+c\bar{d} & i(b\bar{a}-a\bar{b}-d\bar{c}+c\bar{d}) & a\bar{a}-b\bar{b}-c\bar{c}+d\bar{d}\end{pmatrix}.
\end{small}
\end{equation}
We will be primarily concerned with its infinitesimal form. For a particular infinitesimal Lorentz transformation
\begin{equation}
a = \frac{1+\alpha\beta}{1-\gamma}, \ \ b=\alpha, \ \ c=\beta, \ \ d=1-\gamma
\end{equation}
the matrix is given to first order in the infinitesimal parameters by 
\begin{equation}\label{inflorentz}
\Lambda^\mu_\nu = \begin{pmatrix} 1 & \frac{1}{2}(\beta+\bar{\beta}+\alpha+\bar{\alpha}) & \frac{i}{2}(\beta-\bar{\beta}+\bar{\alpha}-\alpha) & -(\gamma+\bar{\gamma})\cr \frac{1}{2}(\alpha+\bar{\alpha}+\bar{\beta}+\beta) & 1 & i(\gamma-\bar{\gamma}) & \frac{1}{2}(\alpha+\bar{\alpha}-\bar{\beta}-\beta)\cr \frac{i}{2}(\bar{\alpha}-\bar{\beta}+\beta-\alpha) & i(\bar{\gamma}-\gamma) & 1 & \frac{i}{2}(\bar{\alpha}+\bar{\beta}-\beta-\alpha)\cr -(\gamma+\bar{\gamma}) & \frac{1}{2}(\beta+\bar{\beta}-\bar{\alpha}-\alpha) & \frac{i}{2}(\beta-\bar{\alpha}-\bar{\beta}+\alpha) & 1\end{pmatrix}.
\end{equation}

\subsection{Review of Massive Spinning Bosons}\label{sec:bosons}
In this subsection we consolidate some useful results from~\cite{Law:2020tsg}. The conformal primary wavefunction for a spin-$s$ massive boson was found to be of the form
\begin{equation}
\label{bosoncpb}
\phi_{\Delta,J,m}^{\mu_1\cdots\mu_s, \pm}(X;\vec{w}) = \int_{\mathcal{H}_3}[d\hat{p}]G_{\Delta+s}\sum_{b=-s}^sg_{J,b}^{(s)}(\vec{w};y,\vec{z})\epsilon_b^{\mu_1\cdots\mu_s}e^{\pm im\hat{p}\cdot X}
\end{equation}
where $J\in\{-s,\cdots, s\}$ is the spin of the conformal primary, $G_{\Delta+s} = (-q\cdot\hat{p})^{-\Delta-s}$ is the scalar bulk to boundary propagator~\cite{Costa:2014kfa}, $\epsilon_b^{\mu_1\cdots\mu_s}$ is the spin-$s$ polarization tensor, the $g_{J,b}^{(s)}$ are scalar functions for each $J,b$ and 
\begin{equation}
\int_{\mathcal{H}_3}[d\hat{p}] \equiv \int_0^\infty \frac{dy}{y^3}\int d^2z = \int\frac{d^3\hat{p}^i}{\hat{p}^0}
\end{equation}
is the integral over a hyperbolic slice. It will be useful to define 
\begin{equation}
\boldsymbol{\epsilon}_J^{\mu_1\cdots\mu_s} \equiv \sum_{b=-s}^s g_{J,b}^{(s)}(\vec{w};y,\vec{z})\epsilon_b^{\mu_1\cdots\mu_s}
\end{equation}
so that~\eqref{bosoncpb} becomes
\begin{equation}
\phi_{\Delta,J,m}^{\mu_1\cdots\mu_s, \pm}(X^\mu;\vec{w}) = \int_{\mathcal{H}_3}[d\hat{p}]G_{\Delta+s}\boldsymbol{\epsilon}_J^{\mu_1\cdots\mu_s}e^{\pm im\hat{p}\cdot X}.
\end{equation}
Comparing the explicit expressions for spin-$1$ and spin-$2$, results in the following expected relations
\begin{eqnarray}\label{spinonespintwo}
\boldsymbol{\epsilon}_{-2}^{\mu\nu}& = & \boldsymbol{\epsilon}_{-1}^\mu\boldsymbol{\epsilon}_{-1}^\nu, \ \ \boldsymbol{\epsilon}_{-1}^{\mu\nu}  =  \frac{i}{\sqrt{2}}(\boldsymbol{\epsilon}_{-1}^\mu\boldsymbol{\epsilon}_0^\nu + \boldsymbol{\epsilon}_0^{\mu}\boldsymbol{\epsilon}_{-1}^\nu), \ \ \boldsymbol{\epsilon}_0^{\mu\nu}=\sqrt{\frac{2}{3}}\left(\boldsymbol{\epsilon}_0^\mu\boldsymbol{\epsilon}_0^\nu - \frac{1}{2}\boldsymbol{\epsilon}_{-1}^\mu\boldsymbol{\epsilon}_{1}^\nu-\frac{1}{2}\boldsymbol{\epsilon}_1^\mu\boldsymbol{\epsilon}_{-1}^\nu\right)\cr
\boldsymbol{\epsilon}^{\mu\nu}_{1} & = & -\frac{i}{\sqrt{2}}(\boldsymbol{\epsilon}_1^\mu\boldsymbol{\epsilon}_0^\nu + \boldsymbol{\epsilon}_0^\mu\boldsymbol{\epsilon}_1^\nu), \ \  \boldsymbol{\epsilon}_2^{\mu\nu} = \boldsymbol{\epsilon}_1^\mu\boldsymbol{\epsilon}_{1}^\nu
\end{eqnarray}
which are consistent with the usual Clebsch-Gordan coefficients up to overall factors resulting from the wavefunction normalization. In a similar fashion, one can construct the higher integer spin fields. It is also useful to note that this is consistent with the construction of the polarization tensors in momentum space~\cite{Hinterbichler:2017qyt}.

\subsection{Dirac Equation}\label{sec:dirac}
In this paper we will consider solutions to the Dirac equation in 4D Minkowski space
\begin{equation}
(\gamma^\mu\partial_\mu-m)\psi = 0.
\end{equation}
The gamma matrices are given by\footnote{We have chosen one particular convention for the gamma matrices. The following construction can be done with any convention choice but will result in differences in overall factors.} 
\begin{equation}
\gamma^0 = \begin{pmatrix} 0 & -\mathbb{I}_2\cr \mathbb{I}_2 & 0\end{pmatrix}, \ \ \gamma^i = \begin{pmatrix} 0 & \sigma^i\cr \sigma^i & 0\end{pmatrix},
\end{equation}
where $\sigma^i$ are the usual Pauli matrices and $\{\gamma^\mu,\gamma^\nu\} = 2\eta^{\mu\nu}\mathbb{I}_4$. We will use familiar conventions where 
\begin{equation}
\boldsymbol{\sigma}  \equiv \{\mathbb{I},\sigma^i\} , \ \ \boldsymbol{\bar{\sigma}} \equiv \{\mathbb{I},-\sigma^i\}.
\end{equation}
Solutions to the Dirac equation are of the form $\psi_s^\pm = u_s^\pm(p)e^{\pm im\hat{p}\cdot X}$ where $\pm$ denote positive and negative energy solutions and $s\in\{\frac{1}{2},-\frac{1}{2}\}$. The spinors satisfy 
\begin{equation}\label{spinoreqn}
(\pm i\gamma^\mu p_\mu-m)u_s^\pm(p) =0.
\end{equation}
The spin representation of the infinitesimal Lorentz transformation~\eqref{inflorentz} can be computed to be 
\begin{equation}
\label{spinlorentz}
\Lambda_{\frac{1}{2}} = \begin{pmatrix} 1-\gamma & \beta & 0 & 0\cr \alpha & 1+\gamma & 0 & 0\cr 0 & 0 & 1+\bar{\gamma} & -\bar{\alpha}\cr 0 & 0 & -\bar{\beta} & 1-\bar{\gamma}\end{pmatrix}.
\end{equation}

\section{Spin-$\frac{1}{2}$ (Dirac) Conformal Primary Wavefunctions}\label{sec:wavefunctions}
In this section we construct the Dirac conformal primary wavefunctions. These wavefunctions are solutions to the massive Dirac equation in 4D as well as 2D conformal primaries. Using symmetry constraints and a convenient ansatz, we are able to find an integral form for the wavefunctions. Once we have the wavefunctions we are able to extract the transformation from a momentum space amplitude to a celestial amplitude.
\subsection{Solving for Wavefunctions Using Constraints}\label{solving}
Inspired by~\cite{Law:2020tsg} we propose the following ansatz for the conformal primary wavefunctions
\begin{equation}
\label{ansatz}
\psi^{\pm}_{\Delta,J,m}(X^\mu;\vec{w}) = \int_{\mathcal{H}_3}[d\hat{p}] \boldsymbol{G}_J^\pm(\hat{p};\vec{w}) e^{\pm im\hat{p}\cdot X} \equiv \int_{\mathcal{H}_3}[d\hat{p}] G_{\Delta+\frac{1}{2}}\boldsymbol{u}_J^\pm(\hat{p};\vec{w}) e^{\pm im\hat{p}\cdot X}
\end{equation}
where $\boldsymbol{u}_J^\pm(\hat{p};\vec{w})$ is a linear combination of the momentum space spinors $u_s^\pm(p)$ and $J \in\{ -\frac{1}{2},\frac{1}{2}\}$ labels the spin\footnote{Much of the previous literature uses spin and helicity interchangeably since for massless fields 2D spin and 4D helicity are the same. For massive fields, spin and helicity are not necessarily the same so we will differentiate between the two when necessary.} of the conformal primary. We consider a linear combination of spins since a general Lorentz transformation transforms a spinor to a linear combination of spinors. The conformal primary wavefunction should transform as a spin-$\frac{1}{2}$ conformal primary and as a spinor under Lorentz transformations. Since the measure and the exponent in~\eqref{ansatz} are invariant under these transformations, we require 
\begin{equation}
\boldsymbol{G}_J(\hat{p};\vec{w})\rightarrow \boldsymbol{G}_J'(\Lambda \hat{p};\vec{w}') \equiv (cw+d)^{\Delta+J}(\bar{c}\bar{w}+\bar{d})^{\Delta-J}\Lambda_{\frac{1}{2}}\boldsymbol{G}_J(\hat{p};\vec{w}).
\end{equation}
Using the coordinate transformations in~\eqref{coordtrans} and the explicit form of $\Lambda_{\frac{1}{2}}$ in~\eqref{spinlorentz} we can expand both sides of this expression to first order in infinitesimal parameters and solve the resulting differential equations. Under such an infinitesimal transformation, we can write an arbitrary function as an expansion
\begin{eqnarray}
f'(z',\bar{z}',y,w',\bar{w}') & = &  f + \alpha(\partial_z+\partial_w)f + \bar{\alpha}(\partial_{\bar{z}}+\partial_{\bar{w}})f \cr
& + & \beta(-z^2\partial_z+y^2\partial_{\bar{z}}-yz\partial_y-w^2\partial_w)f+  \bar{\beta}(y^2\partial_z-\bar{z}^2\partial_{\bar{z}}-y\bar{z}\partial_y-\bar{w}^2\partial_{\bar{w}})f\cr
& + & \gamma(2z\partial_z+2w\partial_w+y\partial_y)f + \bar{\gamma}(2\bar{z}\partial_{\bar{z}}+2\bar{w}\partial_{\bar{w}}+y\partial_y)f.
\end{eqnarray}
Using this to expand $\boldsymbol{G}'_J$ we obtain six matrix-differential equations. Given that we are solving for a function of five variables, this is generally an overconstrained system. However, one can solve these equations, up to overall constants, using the methods outlined in appendix~\ref{app:diffeq}. These constants are then fixed by requiring that the primaries solve the Dirac equation~\eqref{spinoreqn}. For $J=-\frac{1}{2}$ 
\begin{equation}
\psi_{\Delta,-\frac{1}{2},m}^{\pm}(X^\mu;\vec{w}) = \int_{\mathcal{H}_3}[d\hat{p}]G_{\Delta+\frac{1}{2}}\frac{1}{\sqrt{2}}\begin{pmatrix} 1\cr w\cr \mp i\frac{y^2+\bar{z}(z-w)}{y}\cr \mp i\frac{w-z}{y}\end{pmatrix}e^{\pm im\hat{p}\cdot X} 
\end{equation}
and for $J=\frac{1}{2}$
\begin{equation}
\psi_{\Delta,\frac{1}{2},m}^\pm (X^\mu;\vec{w}) = \int_{\mathcal{H}_3}[d\hat{p}]G_{\Delta+\frac{1}{2}}\frac{1}{\sqrt{2}}\begin{pmatrix}\pm i\frac{\bar{z}-\bar{w}}{y}\cr \pm i\frac{y^2+z(\bar{z}-\bar{w})}{y}\cr -\bar{w}\cr 1\end{pmatrix}e^{\pm im\hat{p}\cdot X}.
\end{equation}
To see that these are solutions to the Dirac equation it is important to realize that we can also write them as
\begin{eqnarray}
\psi_{\Delta,-\frac{1}{2},m}^\pm(X^\mu;\vec{w}) & = & \int_{\mathcal{H}_3}[d\hat{p}]G_{\Delta+\frac{1}{2}}\begin{pmatrix} \phi_1(\vec{w})\cr \pm i(\hat{p}\cdot\boldsymbol{\sigma})\phi_1(\vec{w})\end{pmatrix}e^{\pm im\hat{p}\cdot X}\cr
\psi_{\Delta,\frac{1}{2},m}^\pm(X^\mu;\vec{w}) & = & \int_{\mathcal{H}_3}[d\hat{p}]G_{\Delta+\frac{1}{2}}\begin{pmatrix} \mp i(\hat{p}\cdot\boldsymbol{\bar{\sigma}})\phi_2(\vec{w})\cr \phi_2(\vec{w})\end{pmatrix}e^{\pm im\hat{p}\cdot X}
\end{eqnarray}
where $\phi_1(\vec{w}) = \frac{1}{\sqrt{2}}\begin{pmatrix} 1\cr w\end{pmatrix}$ and $\phi_2(\vec{w}) = \frac{1}{\sqrt{2}}\begin{pmatrix} -\bar{w}\cr 1\end{pmatrix}$. We also note that 
\begin{equation}
\phi_1^\dagger(\vec{w}_1)\phi_2(\vec{w}_2) = \overline{\phi_2^\dagger(\vec{w}_2)\phi_1(\vec{w}_1)} = \frac{1}{2}(\bar{w}_1-\bar{w}_2)
\end{equation}
which ensures the  proper orthogonality of the spinors in the integrand as per usual quantum field theory conventions.

\subsection{Amplitude Transformation}\label{amplitudetrans}
One of the primary reasons to study conformal primary wavefunctions is to connect momentum space amplitudes to celestial amplitudes. In the case of massless external particles, for a particular gauge choice, the two are related by a Mellin transform~\cite{Pasterski:2016qvg,Pasterski:2017kqt,Pasterski:2017ylz}
\begin{equation}
\tilde{\mathcal{A}}_n = \prod_{j=1}^n\int_0^\infty d\omega_j \omega_j^{\Delta_j-1}\mathcal{A}_n.
\end{equation}
For massive external scalars the transformation is slightly more complicated and involves a convolution with a scalar bulk to boundary propagator~\cite{Pasterski:2017kqt}
\begin{equation}
\tilde{\mathcal{A}}_n = \prod_{j=1}^n\int_{\mathcal{H}_3}[d\hat{p}_j]G_{\Delta_j}\mathcal{A}_n.
\end{equation}
In the case of massive bosons the transformation was found to be~\cite{Law:2020tsg}
\begin{equation}
\tilde{\mathcal{A}}_{J_1\cdots J_n} = \prod_{j=1}^n\int_{\mathcal{H}_3}[d\hat{p}_j]G_{\Delta_j+s_j}\sum_{b_j=-s_j}^{s_j}g_{J_jb_j}^{(s_j)}\mathcal{A}_{b_1\cdots b_n}
\end{equation}
where the objects in the integral are the same as those that appear in~\eqref{bosoncpb}. This transformation has a more complicated structure since the conformal primary wavefunction involved a sum over polarizations. The amplitude $\mathcal{A}_{b_1\cdots b_n}$ involves $n$ polarization vectors (or tensors) and in transforming to the celestial amplitudes, we have to consider all possible configurations of polarizations, hence the sum in the integrand.

One expects a similar structure for the transformation of fermionic amplitudes. A generic amplitude containing fermions is written in terms of the spinors $u_s$. We are able to write $\boldsymbol{u}_J^\pm$ as a linear combination of the $u_s$ as follows\footnote{There are many conventions for writing down the spinor solutions therefore we have chosen to write this expansion generally to allow the reader to pick their favorite spinors.}
\begin{equation}
\boldsymbol{u}_J^\pm = \left[\frac{\bar{u}^\pm_{-\frac{1}{2}}\boldsymbol{u}^\pm_J}{\bar{u}_{-\frac{1}{2}}^\pm u_{-\frac{1}{2}}^\pm}\right]u^\pm_{-\frac{1}{2}} + \left[\frac{\bar{u}^\pm_{\frac{1}{2}}\boldsymbol{u}^\pm_J}{\bar{u}_{\frac{1}{2}}^\pm u_{\frac{1}{2}}^\pm}\right]u_{\frac{1}{2}}=\sum_{s=-\frac{1}{2}}^{\frac{1}{2}}\mathcal{U}^\pm_{s,J}u_s^\pm.
\end{equation}
These $\mathcal{U}^\pm_{s,J}$ are the fermionic analogs of the $g_{J,b}^{(s)}$ in~\eqref{bosoncpb}. It follows that for an amplitude containing $n$ fermions $\mathcal{A}_{s_1,...,s_n}$ with spin indices $s_i$ the transformation to the conformal primary basis is 
\begin{equation}
\tilde{\mathcal{A}}_{J_1,...,J_n} = \prod_{i=j}^n\int_{\mathcal{H}_3}[d\hat{p}_j]G_{\Delta_j+\frac{1}{2}}\sum_{s_j=-\frac{1}{2}}^{\frac{1}{2}}\mathcal{U}_{s_j,J_j}^\pm\mathcal{A}_{s_1,...,s_n}.
\end{equation}
Usually a scattering amplitude can contain more than one kind of field. For the sake of simplicity we have written the transformation rules for amplitudes containing only one type of particle. In general one will have a product of transformations, one for each type of external particle in the amplitude.

\section{Wavefunction Properties}\label{sec:analytic}
Although it is often more useful to use the integral expressions for the wavefunctions, in some contexts it might be useful to have the precise analytic expression. In this section we write the analytic expression for the wavefunctions by noticing the relation of each component to a scalar conformal primary wavefunction. This allows us to discuss the shadow operators, normalization and massless limit. In order to evaluate the integrals, it is useful to note that the spinor components can be written in terms of scalar conformal primaries of dimension $\Delta+\frac{1}{2}$
\begin{equation}\label{eq:analytic}
\psi_{\Delta,-\frac{1}{2},m}^\pm  =  \frac{1}{\sqrt{2}}\begin{pmatrix} 1\cr w\cr \mp i\left(1+ \frac{\bar{w}\partial_{\bar{w}}}{\Delta-\frac{1}{2}}\right)e^{-\partial_\Delta}\cr \pm\frac{i\partial_{\bar{w}}}{\Delta-\frac{1}{2}}e^{-\partial_\Delta}\end{pmatrix}\phi^\pm_{\Delta+\frac{1}{2}}, \ \ \psi_{\Delta,\frac{1}{2},m}^\pm = \frac{1}{\sqrt{2}}\begin{pmatrix}\pm\frac{i\partial_w}{\Delta-\frac{1}{2}}e^{-\partial_\Delta}\cr \pm i\left(1+\frac{w\partial_w}{\Delta-\frac{1}{2}}\right)e^{-\partial_\Delta}\cr -\bar{w}\cr 1\end{pmatrix}\phi^\pm_{\Delta+\frac{1}{2}}.
\end{equation}
Their analytic structure was determined in~\cite{Pasterski:2017kqt} to be
\begin{equation}
\phi_{\Delta}^\pm(X^\mu;\vec{w}) = \int_{\mathcal{H}_{3}}[d\hat{p}]G_{\Delta}(\hat{p};\vec{w})e^{\pm i m\hat{p}\cdot X} = \frac{4\pi}{im}\frac{(\sqrt{-X^2})^{\Delta-1}}{(-q(\vec{w})\cdot X\mp i\epsilon)^{\Delta}}K_{\Delta-1}(m\sqrt{X^2})
\end{equation}
where $K_\alpha(x)$ is a modified Bessel function of the second kind. Therefore, using the derivative operators in~\eqref{eq:analytic} one can write down a full analytic expression for $\psi_{\Delta,J,m}^\pm$.

\subsection{Shadow Transform}\label{shadow}
A two dimensional conformal primary field $\varphi(w,\bar{w})$ of integer or half integer spin is labelled by weights $h,\bar{h}$ where $\Delta =h+\bar{h}$ and $J=h-\bar{h}$. A shadow field for 2D conformal primaries was defined in~\cite{Osborn:2012vt} to be
\begin{equation}\label{eq:shadow}
\widetilde{\varphi}(w,\bar{w}) = \frac{\Gamma(2-2\bar{h})}{\pi\Gamma(2h-1)}\int d^2z\frac{1}{(w-z)^{2-2h}(\bar{w}-\bar{z})^{2-2\bar{h}}}\varphi(z,\bar{z}).
\end{equation}
The shadow field has weights $(1-h,1-\bar{h})$ therefore conformal dimension $2-\Delta$ and spin $-J$. It was shown in~\cite{Pasterski:2017kqt} that the shadow of a dimension $\Delta$ massive scalar is the same wavefunction with conformal dimension $2-\Delta$ whereas for massless scalars, photons and gravitons the shadow is a different wavefunction though still a conformal primary. For Dirac wavefunctions, the shadow obeys the property 
\begin{equation}
\widetilde{\psi}^\pm_{\Delta,\pm\frac{1}{2},m} = \mp \frac{2i\Gamma(\frac{5}{2}-\Delta)}{\Gamma(\Delta+\frac{1}{2})}\psi_{2-\Delta,\mp\frac{1}{2},m}^\pm.
\end{equation}
The shadows are conformal primaries of dimension $2-\Delta$ and with opposite spin as expected. 

The shadow as defined in~\eqref{eq:shadow} is for spin-$s$ conformal primaries where $2s\in\mathbb{Z}$. This shadow transform is not valid for continuous spin wavefunctions as discussed in~\cite{SimmonsDuffin:2012uy,2018JHEP...11..102K} but is sufficient for our purposes when discussing fermionic fields. It also important to note that our discussion is specifically in 2D. The shadow for a $d$-dimensional conformal primary requires an uplift to the embedding space as discussed for integer spin in~\cite{SimmonsDuffin:2012uy, Pasterski:2017kqt} and for half integer spin in~\cite{Muck:2020wtx}.

\subsection{Dirac Inner Product}\label{diracinnersec}
We demand that the Dirac conformal primaries are delta function normalizable with respect to the Dirac inner product. Such a normalization was also considered in~\cite{Fotopoulos:2020bqj}. We look at the $X_0=0$ Cauchy slice where this inner product is given by
\begin{equation}\label{diracinner}
(\psi_{\Delta_1,J,m},\psi_{\Delta_2,J',m}) = \frac{1}{2}\int d^3X (\bar{\psi}_{\Delta_1,J}\gamma^0\psi_{\Delta_2,J'}+\bar{\psi}_{\Delta_2,J'}\gamma^0\psi_{\Delta_1,J}).
\end{equation}
To compute this inner product we note that the spinors have the property
\begin{eqnarray}
\boldsymbol{\bar{u}}_J^\pm(\vec{w}_1)\gamma^0 \boldsymbol{u}_{J'}^\pm(\vec{w}_2) & = &   -\hat{p}^0 \left[\frac{y^2+(\bar{w}_1-\bar{z})(w_2-z)}{y}\delta_{J+J',-1}+ \frac{y^2+(w_1-z)(\bar{w}_2-\bar{z})}{y}\delta_{J+J',1}\right]\cr
& \mp & i\hat{p}^0\left[(w_1-w_2)\delta_{J-J',1}+(\bar{w}_1-\bar{w}_2)\delta_{J-J',-1}\right]
\end{eqnarray}
and that the spatial integral is 
\begin{equation}
\int d^3Xe^{i(p-p')\cdot X} = \frac{2(2\pi)^3y^3}{m^2\hat{p}^0}\delta^{(2)}(z-z')\delta(y-y') = \frac{(2\pi)^3}{m^2}\delta^{(3)}(\hat{p}-\hat{p}').
\end{equation}
where $\delta^{(3)}(\hat{p}-\hat{p}')$ is the Lorentz invariant delta function on $\mathcal{H}_3$. 
Using the simplifications above and the results in appendix~\ref{app:bulkint}, the inner product is
\begin{eqnarray}
(\psi_{1+i\lambda_1,J,m}^\pm&,&\psi_{1+i\lambda_2,J',m}^\pm)  =  \mp \frac{2(2\pi)^6}{m^2(1+4\lambda_1^2)}\delta_{JJ'}\delta(\lambda_1-\lambda_2)\delta^{(2)}(w_1-w_2)\cr
 &\pm & \frac{i(2\pi)^5\delta(\lambda_1+\lambda_2)}{m^2}\left[\frac{\delta_{J-J',1}\partial_{\bar{w}_1}+\delta_{J-J',-1}\partial_{w_1}}{(\frac{1}{2}-i\lambda_1)^2}\frac{1}{|w_{12}|^{2(\frac{1}{2}-i\lambda_1)}}-\mbox{c.c}\right].
\end{eqnarray}
Just as in~\cite{Pasterski:2017kqt}, convergence of the inner product requires that the wavefunctions be on the principal series where $\Delta=1+i\lambda$ for $\lambda\in\mathbb{R}$. We see that the wavefunctions are only delta function normalizable if $J=J'$. This gives us two ways to make a basis in order to remove redundancies. We can choose the basis elements to have both spins, $J=\pm \frac{1}{2}$, in which case we need to restrict $\lambda\in\mathbb{R}_{+\cup 0}$. The wavefunctions with $\lambda\in\mathbb{R}_{-\cup 0}$ will be related by a shadow transform. Alternatively, we can choose the basis elements to have just one spin $J=\frac{1}{2}$ or $J=-\frac{1}{2}$ and let $\lambda\in\mathbb{R}$. Here, the wavefunctions with opposite spin will be related by a shadow transform. In what follows we choose the latter.

Though we make a particular choice of basis, it is important to note how the two choices are related to one another. If we start with the set $\{J=\frac{1}{2},\lambda\in\mathbb{R}\}$ we can write it as the union 
\begin{equation}
\left\{J=\frac{1}{2},\lambda\in\mathbb{R}\right\} = \left\{J=\frac{1}{2},\lambda\in\mathbb{R}_{-\cup 0}\right\}\cup\left\{J=\frac{1}{2},\lambda\in\mathbb{R}_{+\cup 0}\right\}.
\end{equation}
We can take the first set and take the shadow of those wavefunctions to get 
\begin{equation}
\mbox{Shadow}\left(\left\{J=\frac{1}{2},\lambda\in\mathbb{R}_{-\cup 0}\right\}\right)=\left\{J=-\frac{1}{2},\lambda\in\mathbb{R}_{+\cup 0}\right\}.
\end{equation}
Therefore we see that 
\begin{equation}
\mbox{Shadow}\left(\left\{J=\frac{1}{2},\lambda\in\mathbb{R}_{-\cup 0}\right\}\right)\cup\left\{J=\frac{1}{2},\lambda\in\mathbb{R}_{+\cup 0}\right\} = \left\{J=\pm\frac{1}{2},\lambda\in\mathbb{R}_{+\cup 0}\right\}
\end{equation}
which is the other choice of basis.

\subsection{Completeness}\label{completesec}
We can now show that our choice of basis is complete. One can show for the spinor solutions that
\begin{equation}
\mp 2i\boldsymbol{\bar{u}}_J^\pm(\hat{p}_1;\vec{w}) \boldsymbol{u}_J^\pm(\hat{p}_2;\vec{w}) = -(\hat{p}_1+\hat{p}_2)\cdot \hat{q}(\vec{w}).
\end{equation}
In appendix~\ref{app:boundary} we show that\footnote{We would like to thank Monica Pate and Ana-Maria Raclariu for help with the details of this calculation.} 
\begin{equation}\label{complete}
\int_{-\infty}^\infty d\lambda \mu(\lambda) \int d^2w\left[G_{\frac{3}{2}-i\lambda}(\hat{p}_1;\vec{w})G_{\frac{1}{2}+i\lambda}(\hat{p}_2;\vec{w}) + G_{\frac{1}{2}-i\lambda}(\hat{p}_1;\vec{w})G_{\frac{3}{2}+i\lambda}(\hat{p}_2;\vec{w})\right]=\delta^{(3)}(\hat{p}_1,\hat{p}_2)
\end{equation}
for 
\begin{equation}
\mu(\lambda) = \frac{2\Gamma(\frac{3}{2}+i\lambda)\Gamma(\frac{3}{2}-i\lambda)}{\Gamma(\frac{1}{2}+i\lambda)\Gamma(\frac{1}{2}-i\lambda)}.
\end{equation}
Using this, one finds that \begin{equation}
e^{\pm im\hat{p}\cdot X}= \mp i\int_{-\infty}^\infty d\lambda\frac{4\Gamma(\frac{3}{2}+i\lambda)\Gamma(\frac{3}{2}-i\lambda)}{\Gamma(\frac{1}{2}+i\lambda)\Gamma(\frac{1}{2}-i\lambda)}\int d^2wG_{\frac{3}{2}-i\lambda}(\hat{p};\vec{w})\boldsymbol{\bar{u}}^\pm_J\psi_{1+i\lambda,J,m}^\pm(X^\mu;\vec{w}).
\end{equation}
Therefore, the transformation to celestial amplitudes is invertible and the basis is complete. Since we can transform from one basis choice to another using the shadow, we know the other basis choice is also complete.

\subsection{Massless Limit}\label{masslesssec}
One can also take the massless limit of the massive wavefunctions. In the massless limit~\cite{Pasterski:2017kqt} the bulk to boundary propagator is 
\begin{equation}
G_{\Delta}(y,\vec{z};\vec{w})\rightarrow \frac{\pi\Gamma(\Delta-1)}{\Gamma(\Delta)}y^{2-\Delta}\delta^{(2)}(z-w).
\end{equation}
Substituting this and using the change of integration variable $\omega = \frac{m}{2y}$, the massless limit of the spin-$\frac{1}{2}$ wavefunctions is 
\begin{equation}
\psi_{\Delta,-\frac{1}{2}}^\pm(X^\mu;\vec{w}) = \int_0^\infty d\omega \omega^{\Delta-\frac{1}{2}}\begin{pmatrix}\phi_1\cr 0\end{pmatrix}e^{\pm i q\cdot X}, \ \ \psi_{\Delta,\frac{1}{2}}^\pm(X^\mu;\vec{w}) = \int_0^\infty d\omega \omega^{\Delta-\frac{1}{2}}\begin{pmatrix} 0\cr \phi_2\end{pmatrix}e^{\pm iq\cdot X}
\end{equation}
This is consistent with the massless wavefunctions in~\cite{Fotopoulos:2020bqj}, where they were used to study supersymmetry on the CCFT, up to differences between our conventions for the Dirac equation and the conventions in~\cite{Taylor:2017sph}.

\section{Momentum Generators}\label{sec:momentum}
It was proven useful to constrain celestial amplitudes using Poincar\'e symmetries in~\cite{Law:2019glh,Stieberger:2018onx}. In the massless case, the momentum generators are diagonal and do not mix the spin, however it was found in the massive bosonic case that they are not diagonal. Rather, they mix neighboring spins. In the fermionic case, the eigenvalue equation one wishes to solve is 
\begin{equation}
\sum_{I=-\frac{1}{2}}^{\frac{1}{2}}P^{\mu,\pm}_{J,I}\boldsymbol{G}_J^\pm =m\hat{p}^\mu \boldsymbol{G}_J^\pm.
\end{equation}
We are able to find that the momentum generator is 
\begin{small}
\begin{eqnarray}
& & P^{\mu,\pm}_{J,I} =  \frac{m}{2}\bigg[\left[\frac{\pm i\delta_{J,I+1}}{\Delta-\frac{3}{2}}\left(\partial_w q^\mu + \frac{q^\mu\partial_w}{\Delta-\frac{1}{2}}\right)+\frac{\pm i\delta_{J+1,I}}{\Delta-\frac{3}{2}}\left(\partial_{\bar{w}}q^\mu+\frac{q^\mu\partial_{\bar{w}}}{\Delta-\frac{1}{2}}\right)\right]\cr
& + & \delta_{J,I}\left[\left[\partial_w\partial_{\bar{w}}q^\mu + \frac{(\partial_w q^\mu)\partial_{\bar{w}}}{\Delta-1-J}+\frac{(\partial_{\bar{w}}q^\mu)\partial_w}{\Delta-J}+\frac{q^\mu\partial_w\partial_{\bar{w}}}{(\Delta-\frac{3}{2})(\Delta-\frac{1}{2})}\right]e^{-\partial_\Delta} + \frac{(\Delta+\frac{1}{2})q^\mu}{\Delta-\frac{1}{2}}e^{\partial_\Delta}\right]\bigg]
\end{eqnarray}\end{small}Up to normalization, this is the same result as (3.3) in~\cite{Law:2020tsg} with $s=\frac{1}{2}$. 
Therefore, one can show that this operator properly squares to $-m^2$ and that along with the Lorentz generators forms a representation of the Poincar\'e algebra. We do not reproduce it here.

The momentum generator is not diagonal in this representation because, just as in the bosonic case, the wavefunction is constructed with a sum over all spins. The fact that the momentum generator for massive fermionic fields is structurally the same as that of the massive bosonic fields is not surprising given the relation between the fields shown in the next section.

\section{Arbitrary Integer and Half-Integer Spin}\label{sec:spin1}
In this section we first relate the spinor solutions to the spin-1 polarization vectors. We are then able to extrapolate these results to outline a method of writing down the massive conformal primary wavefunctions for arbitrary integer and half-integer spin\footnote{It is useful to note that it might be even cleaner to express these relations using the massive spinor-helicity formalism~\cite{Arkani-Hamed:2017jhn} however we have not explored that in this paper.}.

\subsection{Relation to Spin-$1$ Polarizations}\label{spinone}
In this section we relate the spinors $\boldsymbol{u}_J^\pm$ to the linear combination of polarization vectors that appear in the massive spin-1 conformal primary wavefunctions in~\cite{Law:2020tsg}. Arbitrary spin bosonic polarization tensors can be constructed from the spin-1 polarization vectors by taking appropriate linear combinations of tensor products~\cite{Hinterbichler:2017qyt} e.g $\epsilon^{\mu\nu}_2\propto \epsilon^\mu_1\epsilon^\nu_1$. The relation between spin-1 and spin-2 was outlined in~\eqref{spinonespintwo}. To construct these polarization vectors from the Dirac spinors it is necessary to consider a product that includes a gamma matrix to get the appropriate index structure. We consider the inner product $\boldsymbol{\bar{u}}_J^\pm\gamma^\mu \boldsymbol{u}_{J'}^\pm$. We can relate them to the $g_{J,b}$ appearing in~\eqref{bosoncpb}
\begin{equation}
\bar{\boldsymbol{u}}_{\pm\frac{1}{2}}^\mp\gamma^\mu\boldsymbol{u}_{\pm\frac{1}{2}}^\pm =  -\boldsymbol{\epsilon}^\mu_0, \ \ \ \
\bar{\boldsymbol{u}}_{-\frac{1}{2}}^\pm\gamma^\mu \boldsymbol{u}_{\frac{1}{2}}^\mp  =  \pm i\boldsymbol{\epsilon}_1^\mu, \ \ \ \ 
\bar{\boldsymbol{u}}_{\frac{1}{2}}^\pm\gamma^\mu \boldsymbol{u}_{-\frac{1}{2}}^\mp  = \mp i\boldsymbol{\epsilon}_{-1}^\mu.
\end{equation}
Therefore we explicitly have the relations
\begin{eqnarray}
\phi_{\Delta,J=-1,m}^{\mu, \pm} & = & \mp i\int_{\mathcal{H}_3}[d\hat{p}]G_{\Delta+1}\left[\bar{\boldsymbol{u}}_{\frac{1}{2}}^{\mp}\gamma^\mu\boldsymbol{u}_{-\frac{1}{2}}^{\pm}\right]e^{\pm im\hat{p}\cdot X}\cr
\phi_{\Delta,J=0,m}^{\mu, \pm} & = & -\frac{1}{2}\int_{\mathcal{H}_3}[d\hat{p}]G_{\Delta+1}\left[\bar{\boldsymbol{u}}^{\mp}_{-\frac{1}{2}}\gamma^\mu\boldsymbol{u}_{-\frac{1}{2}}^{\pm} + \bar{\boldsymbol{u}}^\mp_{\frac{1}{2}}\gamma^\mu\boldsymbol{u}_{\frac{1}{2}}^\pm\right]e^{\pm im\hat{p}\cdot X}\cr
\phi_{\Delta,J=1,m}^{\mu, \pm} & = & \pm i\int_{\mathcal{H}_3}[d\hat{p}]G_{\Delta+1}\left[\bar{\boldsymbol{u}}_{-\frac{1}{2}}^{\mp}\gamma^\mu\boldsymbol{u}_{\frac{1}{2}}^{\pm}\right]e^{\pm im\hat{p}\cdot X}.
\end{eqnarray}
Whereas it may appear odd that we need to pair positive and negative energy spinor solutions to construct the spin-$1$ wavefunctions, we are just matching helicities. For example, the positive energy $||s|=1,J=-1\rangle$ wavefunction has negative helicity therefore we obtain it from a combination of the negative helicity spinors: positive energy $||s|=\frac{1}{2},J=-\frac{1}{2}\rangle$ and negative energy $||s|=\frac{1}{2},J=\frac{1}{2}\rangle$.

\subsection{Rarita-Schwinger Fields}\label{rarita}
We have successfully shown how to construct the massive spin-1 conformal primary wavefunction from the Dirac spinors. It is rather convenient, and expected, that they are related by the usual Clebsch-Gordan coefficients. Since we already know how to construct higher integer spin wavefunctions, the natural question is whether we can construct higher half-integer spin wavefunctions. To that end, we can construct massive spin-$\frac{3}{2}$ conformal primary wavefunctions, the Rarita-Schwinger fields. Constructed this way, a spin-$\frac{3}{2}$ massive conformal primary wavefunction can be written\footnote{In principle one can derive these wavefunctions using the same methods in subsection~\ref{solving} however as the spin increases, the equations are more difficult to solve. It is easy to see that by construction, these wavefunctions satisfy the Lorentz and conformal transformation properties.}
\begin{equation}
\Psi_{\Delta,J,m}^{\mu,\pm}(X^\mu;\vec{w}) = \int_{\mathcal{H}_3}[d\hat{p}]G_{\Delta+\frac{3}{2}}\boldsymbol{u}_J^{\mu,\pm}(p;\vec{w})e^{\pm im\hat{p}\cdot X}
\end{equation}
where $J\in\{-\frac{3}{2},-\frac{1}{2},\frac{1}{2},\frac{3}{2}\}$. The integration functions here are given by 
\begin{eqnarray}
\boldsymbol{u}_{\frac{3}{2}}^{\mu,\pm} = \boldsymbol{\epsilon}_1^\mu\boldsymbol{u}_{\frac{1}{2}}^\pm, & \ \ &  \boldsymbol{u}_{\frac{1}{2}}^{\mu,\pm} = \sqrt{\frac{1}{3}}\boldsymbol{\epsilon}_1^\mu\boldsymbol{u}_{-\frac{1}{2}}^\pm + i\sqrt{\frac{2}{3}}\boldsymbol{\epsilon}_0^\mu\boldsymbol{u}_{\frac{1}{2}}^\pm \cr
\boldsymbol{u}_{-\frac{1}{2}}^{\mu,\pm} = \sqrt{\frac{1}{3}}\boldsymbol{\epsilon}_{-1}^\mu\boldsymbol{u}_{\frac{1}{2}}^\pm + i\sqrt{\frac{2}{3}}\boldsymbol{\epsilon}_0^\mu\boldsymbol{u}_{-\frac{1}{2}}^\pm, & \ \ & \boldsymbol{u}_{-\frac{3}{2}}^{\mu,\pm} = \boldsymbol{\epsilon}_{-1}^\mu\boldsymbol{u}_{-\frac{1}{2}}^\pm
\end{eqnarray}
up to overall constants that can be chosen to fit an appropriate normalization scheme. It is also useful to note that these relations hold true in the massless case so the massless limit of these wavefunctions is consistent with the massless spin-$\frac{3}{2}$ fields in~\cite{Fotopoulos:2020bqj}. In this way one can construct any massive integer or half integer spin wavefunction.

Upon constructing higher half integer spin wavefunctions one should be able to show they are normalizable with respect to an appropriately defined inner product. One would expect that they would be constrained to be on the principal series and that the basis would be complete. We have chosen not to generalize these results with the expectation that it can be done if necessary.  

\section*{Acknowledgements}
I am grateful to Alex Atanasov, Erin Crawley, Alfredo Guevara, Rajamani Narayanan, Aditya Parikh, Monica Pate, Ana-Maria Raclariu, Andrew Strominger, and Evan Zayas for many useful discussions and important insights. This work was supported by DOE grant de-sc/0007870.

\appendix
\section{Solving Constraint Equations}\label{app:diffeq}
In this appendix we discuss a particular caveat in finding the solutions to the differential equations in section~\ref{sec:wavefunctions}. In particular there are six differential equations of the form
\begin{equation}
\mathcal{D}_i \boldsymbol{u}_J = \mathcal{A}_i\boldsymbol{u}_J
\end{equation}
where the $\mathcal{D}_i$ are differential operators for $i=1,\cdots, 6$ given by 
\begin{eqnarray}
\mathcal{D}_1  =  \partial_w+\partial_z, & \ \ & \mathcal{D}_2 = \partial_{\bar{w}}+\partial_{\bar{z}}\cr
\mathcal{D}_3 = -(w^2\partial_w + yz\partial_y+z^2\partial_z-y^2\partial_{\bar{z}}) & \ \ & \mathcal{D}_4 = -(\bar{w}^2\partial_{\bar{w}}+y\bar{z}\partial_y - y^2\partial_z+\bar{z}^2\partial_{\bar{z}})\cr
\mathcal{D}_5 = 2w\partial_w +y\partial_y + 2z\partial_z & \ \ & \mathcal{D}_6 = 2\bar{w}\partial_{\bar{w}}+y\partial_y + 2\bar{z}\partial_{\bar{z}}.
\end{eqnarray}
There are two different linear combinations of different subsets of the set of $\mathcal{D}_i$ that are equal to $\partial_y$. In particular 
\begin{eqnarray}
\partial_y & = & \frac{(\bar{z}-\bar{w})}{y(y^2+|z-w|^2)}\left(2wz\mathcal{D}_1-2\mathcal{D}_3-(w+z)\mathcal{D}_5 +\frac{y^2(\mathcal{D}_6-2\bar{w}\mathcal{D}_2)}{\bar{z}-\bar{w}}\right)\cr
\partial_y & = &  \frac{(z-w)}{y(y^2+|z-w|^2)}\left(2\bar{w}\bar{z}\mathcal{D}_2-2\mathcal{D}_4-(\bar{w}+\bar{z})\mathcal{D}_6+\frac{y^2(\mathcal{D}_5-2w\mathcal{D}_1)}{z-w}\right).
\end{eqnarray}
This results in two differential equations
\begin{equation}
\partial_y \boldsymbol{u}_J = \mathcal{M}_1\boldsymbol{u}_J, \ \ \ \partial_y \boldsymbol{u}_J = \mathcal{M}_2\boldsymbol{u}_J
\end{equation}
where $\mathcal{M}_1,\mathcal{M}_2$ are two $4\times 4$ matrices. These are just differential equations for the $y$ dependence of the $\boldsymbol{u}_J$ therefore we can let $\boldsymbol{u}_J = f_J(\vec{z},\vec{w})\boldsymbol{\tilde{u}}_J$. Setting the two equations equal to one another, and focusing on just the $y$ dependence one notices that the $\boldsymbol{\tilde{u}}_J$ must satisfy
\begin{equation}
(\mathcal{M}_1-\mathcal{M}_2)\boldsymbol{\tilde{u}}_J=0.
\end{equation}
 This is solved assuming that the nullspace of $\mathcal{M}_1-\mathcal{M}_2$ is non-empty. In this particular case, there are two independent vectors in the nullspace for each spin
\begin{equation}
\boldsymbol{\tilde{u}}_{-\frac{1}{2}}  =  \begin{cases} \begin{pmatrix} 0 \cr 0\cr\frac{1}{w}\cr 1\end{pmatrix}, &  \begin{pmatrix} \frac{y^2}{w-z}-\bar{z} \cr 1\cr 0\cr 0\end{pmatrix}\end{cases}, \ \ \boldsymbol{\tilde{u}}_{\frac{1}{2}}  = \begin{cases} \begin{pmatrix} 0\cr 0\cr \frac{\bar{z}-\bar{w}}{y^2+z(\bar{z}-\bar{w})}\cr 1\end{pmatrix}, & \begin{pmatrix} -\bar{w}\cr 1\cr 0\cr 0\end{pmatrix}\end{cases}
\end{equation}
 Therefore we are able to write $\boldsymbol{u}_J$ as linear combination of the two vectors in the nullspace and solve for the coefficients, $f_J(\bar{z},\bar{w})$ by asserting that the initial six differential equations must be satisfied.

\section{Bulk Integrals of Propagators}\label{app:bulkint}
In this appendix we include details required for the computation of the integrals appearing in the Dirac inner product. Similar integrals have been discussed in~\cite{Costa:2014kfa}. We would like to compute an integral of the form
\begin{equation}\label{bulkintegral}
\mathcal{I}_{\rm{bulk}} = \mathcal{I}_1 +\mathcal{I}_2 = \int_0^\infty \frac{dy}{y^3}\int d^2z G_{\alpha+\frac{1}{2}-i\lambda_1}G_{\alpha+\frac{1}{2}+i\lambda_2}\left[y+\frac{(\bar{w}_1-\bar{z})(w_2-z)}{y}\right].
\end{equation}
This integral has two terms which are to be computed separately. Both can be computed using the Fourier transform\footnote{We have defined $k=k_1-ik_2$ so that the terms look nice.}
\begin{equation}\label{Fourier}
\frac{1}{(y^2+|z-w_i|^2)^{\Delta}} \equiv \int \frac{d^2k}{(2\pi)^2}\int d^2x \frac{e^{ik\cdot (x-(z-w_i))}}{(y^2+x^2)^{\Delta}} = \int \frac{dkd\bar{k}}{(2\pi)^2}\int dxd\bar{x}\frac{e^{ikx+i\bar{k}\bar{x}}}{(y^2+x\bar{x})^{\Delta}}.
\end{equation}
The first term in~\eqref{bulkintegral} is 
\begin{small}
\begin{eqnarray}
\mathcal{I}_1  & = & \int_0^\infty \frac{dy}{y^2}\int d^2z\left(\frac{y}{y^2+|z-w_1|^2}\right)^{\alpha+\frac{1}{2}-i\lambda_1}\left(\frac{y}{y^2+|z-w_2|^2}\right)^{\alpha+\frac{1}{2}+i\lambda_2}\cr
& = & \int_0^\infty dy y^{2\alpha-1-i(\lambda_1-\lambda_2)}\int d^2z\int \frac{d^2k_1d^2k_2}{(2\pi)^4}\int d^2x_1d^2x_2 \frac{e^{ik_1\cdot(x_1-(z-w_1))+ik_2\cdot(x_2-(z-w_2))}}{(y^2+x_1^2)^{\alpha+\frac{1}{2}-i\lambda_1}(y^2+x_2^2)^{\alpha+\frac{1}{2}+i\lambda_2}}\cr
& = & 2\int_0^\infty dy y^{2\alpha-1-i(\lambda_1-\lambda_2)}\int \frac{d^2k}{(2\pi)^2}\int d^2x_1d^2x_2\frac{e^{ik\cdot(x_1-x_2)+ik\cdot(w_1-w_2)}}{(y^2+x_1^2)^{\alpha+\frac{1}{2}-i\lambda_1}(y^2+x_2^2)^{\alpha+\frac{1}{2}+i\lambda_2}}
\end{eqnarray}\end{small}where to go to the last line we have integrated over $z,\bar{z}$ to get delta functions in the $k_i$ and then relabeled $k_1\rightarrow k$. Next we use Schwinger parameters to write 
\begin{equation}
\frac{1}{(y^2+x^2)^\Delta} = \frac{1}{\Gamma(\Delta)}\int_0^\infty d\beta \beta^{\Delta-1}e^{-\beta(y^2+x^2)}
\end{equation}
for the two factors appearing the integrand. Then, we can complete the square in the exponent to easily perform the $x_i$ integrals. This leaves us with 
\begin{eqnarray}
\mathcal{I}_1 & = &  \frac{2\pi^2}{\Gamma(\alpha+\frac{1}{2}-i\lambda_1)\Gamma(\alpha+\frac{1}{2}+i\lambda_2)}\int_0^\infty dy y^{2\alpha-1-i(\lambda_1-\lambda_2)}\int\frac{d^2k}{(2\pi)^2}e^{ik\cdot(w_1-w_2)}\cr
& \times & \int_0^\infty d\beta_1 d\beta_2 \beta_1^{\alpha-\frac{3}{2}-i\lambda_1}\beta_2^{\alpha-\frac{3}{2}+i\lambda_2}e^{-\beta_1y^2-\beta_2 y^2-\frac{k^2}{4\beta_1}-\frac{k^2}{4\beta_2}}.
\end{eqnarray}
Next we rescale $\beta_i\rightarrow \frac{|\vec{k}|\beta_i}{2y}$ then $y\rightarrow \frac{2y}{|\vec{k}|}$ and perform the integral over $y$ to get
\begin{eqnarray}\label{term1}
\mathcal{I}_1 & = & \frac{2^{3-2\alpha+i(\lambda_1-\lambda_2)}\pi^2}{\Gamma(\alpha+\frac{1}{2}-i\lambda_1)\Gamma(\alpha+\frac{1}{2}+i\lambda_2)} \int\frac{d^2k}{(2\pi)^2}e^{ik\cdot(w_1-w_2)}|\vec{k}|^{2\alpha-2-i(\lambda_1-\lambda_2)} \cr
& \times & \int_0^\infty d\beta_1 d\beta_2 \frac{\beta_1^{\alpha-\frac{3}{2}-i\lambda_1}\beta_2^{\alpha-\frac{3}{2}+i\lambda_2}}{(\beta_1+\beta_2+\frac{1}{\beta_1}+\frac{1}{\beta_2})}.
\end{eqnarray}
Before evaluating the remaining integrals, it is useful to evaluate the second term in~\eqref{bulkintegral}
\begin{small}
\begin{eqnarray}
\mathcal{I}_2 & = &  \int_0^\infty \frac{dy}{y^4}\int d^2z(\bar{w}_1-\bar{z})(w_2-z)\left(\frac{y}{y^2+|z-w_1|^2}\right)^{\alpha+\frac{1}{2}-i\lambda_1}\left(\frac{y}{y^2+|z-w_2|^2}\right)^{\alpha+\frac{1}{2}+i\lambda_2}\cr
& = & 2\int_0^\infty dy y^{2\alpha-3-i(\lambda_1-\lambda_2)}\int\frac{d^2k}{(2\pi)^2}\int d^2x_1 d^2x_2\frac{\bar{x}_1 x_2e^{ik\cdot(x_1-x_2)+ik\cdot(w_1-w_2)}}{(y^2+x_1^2)^{\alpha+\frac{1}{2}-i\lambda_1}(y^2+x_2^2)^{\alpha+\frac{1}{2}+i\lambda_2}}
\end{eqnarray}\end{small}which can be done in a similar way using Schwinger parameters to get
\begin{eqnarray}
\mathcal{I}_2 & = & \frac{ 2^{3-2\alpha+i(\lambda_1-\lambda_2)}\pi^2}{\Gamma(\alpha+\frac{1}{2}-i\lambda_1)\Gamma(\alpha+\frac{1}{2}+i\lambda_2)}\int\frac{d^2k}{(2\pi)^2}e^{ik\cdot(w_1-w_2)}|\vec{k}|^{2\alpha-2-i(\lambda_1-\lambda_2)}\cr
& \times & \int_0^\infty d\beta_1 d\beta_2\frac{\beta_1^{\alpha-\frac{5}{2}-i\lambda_1}\beta_2^{\alpha-\frac{5}{2}+i\lambda_2}}{(\beta_1+\beta_2+\frac{1}{\beta_1}+\frac{1}{\beta_2})}.
\end{eqnarray}
Combining this with the first term~\eqref{term1} and performing the coordinate transformation $\beta_1={e^{U+V}}$ and $\beta_2 = e^{U-V}$ we obtain
\begin{eqnarray}
\mathcal{I}_{\rm{bulk}}& = & \frac{2^{4-2\alpha+i(\lambda_1-\lambda_2)}\pi^2}{\Gamma(\alpha+\frac{1}{2}-i\lambda_1)\Gamma(\alpha+\frac{1}{2}+i\lambda_2)}\int\frac{d^2k}{(2\pi)^2}e^{ik\cdot(w_1-w_2)}|\vec{k}|^{2\alpha-2-i(\lambda_1-\lambda_2)}\cr
& \times & \int_{-\infty}^\infty dU dV\frac{e^{(2\alpha-2-i(\lambda_1-\lambda_2))U}e^{-i(\lambda_1+\lambda_2)V}}{2\cosh V}.
\end{eqnarray}
When $\alpha=1$ both exponentials become purely imaginary and we can evaluate this exactly
\begin{equation}
\frac{4\pi^3\delta(\lambda_1-\lambda_2)\delta^{(2)}(w_1-w_2)}{\Gamma(\frac{3}{2}-i\lambda_1)\Gamma(\frac{3}{2}+i\lambda_2)}\int_{-\infty}^\infty dV\frac{e^{-2i\lambda_1V}}{\cosh V}.
\end{equation}
In order to evaluate the entire inner product, we need to consider the second term in~\eqref{diracinner} which is the complex conjugate of this term. Combining the two gives us 
\begin{equation}
\frac{8\pi^3\delta(\lambda_1-\lambda_2)\delta^{(2)}(w_1-w_2)}{\Gamma(\frac{3}{2}+i\lambda_1)\Gamma(\frac{3}{2}-i\lambda_1)}\int_{-\infty}^\infty dV\frac{\cos(2\lambda_1 V)}{\cosh V} = \frac{32\pi^3\delta(\lambda_1-\lambda_2)\delta^{(2)}(w_1-w_2)}{1+4\lambda_1^2}.
\end{equation}

\section{Boundary Integrals of Propagators}\label{app:boundary}
In this appendix we include details required for the computation of the integrals appearing in the completeness relation. Similar integrals have been discussed in~\cite{Costa:2014kfa,SimmonsDuffin:2012uy,monicanote} but for different pairs of conformal weights. We would like to compute an integral of the form 
\begin{equation}
\mathcal{I}_1  =  \int d^2w\left(\frac{y_1}{y_1^2+|z_1-w|^2}\right)^{\frac{3}{2}-i\lambda}\left(\frac{y_2}{y_2^2+|z_2-w|^2}\right)^{\frac{1}{2}+i\lambda}.
\end{equation}
We can use the Fourier representation in~\eqref{Fourier}, integrate over $w,\bar{w}$ in the same way as in the bulk case and rescale integration parameters $\beta_1,\beta_2$ to obtain
\begin{eqnarray}
\mathcal{I}_1 & = & y_1^{\frac{3}{2}-i\lambda}y_2^{\frac{1}{2}+i\lambda}\int d^2w\int\frac{d^2k_1 d^2k_2}{(2\pi)^4}\int d^2x_1 d^2x_2\frac{e^{ik_1\cdot(x_1-(z_1-w))}e^{ik_2\cdot(x_2-(z_2-w))}}{(y_1^2+x_1^2)^{\frac{3}{2}-i\lambda}(y_2^2+x_2^2)^{\frac{1}{2}+i\lambda}}\cr
& = & \frac{2\pi^2y_1y_2}{\Gamma(\frac{3}{2}-i\lambda)\Gamma(\frac{1}{2}+i\lambda)}\int\frac{d^2k}{(2\pi)^2}e^{ik\cdot(z_1-z_2)}\cr
& \times & \int_0^\infty d\beta_1 d\beta_2 \beta_1^{-\frac{1}{2}-i\lambda}\beta_2^{-\frac{3}{2}+i\lambda}e^{-\frac{|\vec{k}|\beta_1 y_1}{2}-\frac{|\vec{k}|\beta_2 y_2}{2}-\frac{|\vec{k}|y_1}{2\beta_1}-\frac{|\vec{k}|y_2}{2\beta_2}}.
\end{eqnarray}
Using 3.471.9 in~\cite{gradshteyn2007} we can perform the $\beta_i$ integrals to get Bessel functions
\begin{equation}
\mathcal{I}_1  =  \frac{8\pi^2 y_1y_2}{\Gamma(\frac{3}{2}-i\lambda)\Gamma(\frac{1}{2}+i\lambda)}\int\frac{d^2k}{(2\pi)^2}e^{ik\cdot(z_1-z_2)}K_{\frac{1}{2}-i\lambda}(|\vec{k}|y_1)K_{-\frac{1}{2}+i\lambda}(|\vec{k}|y_2).
\end{equation}
Using the representations of the modified Bessel function of the second kind
\begin{equation}
K_\nu(x) =  \frac{1}{\cos\frac{\nu\pi}{2}}\int_0^\infty \cos(x\sinh t)\cosh\nu t dt =  \frac{1}{\sin\frac{\nu\pi}{2}}\int_0^\infty \sin(x\sinh t)\sinh \nu t dt
\end{equation}
which is valid when $-1<\mbox{Re}(\nu)<1$, 
\begin{eqnarray}\label{integralone}
\mathcal{I}_1 & = & 16\pi y_1y_2\int\frac{d^2k}{(2\pi)^2}e^{ik\cdot z_{12}}\int_0^\infty du dt\cos(|\vec{k}|y_1\sinh u)\sin(|\vec{k}|y_2\sinh t)\cr
& \times & \frac{\frac{1}{2}+i\lambda}{\frac{1}{4}+\lambda^2}\cosh\left[\left(\frac{1}{2}-i\lambda\right)u\right]\sinh\left[\left(\frac{1}{2}-i\lambda\right)t\right].
\end{eqnarray}
To obtain the appropriate completeness relation, we need to consider the other term in~\ref{complete} which can be identified with this term where $\lambda\rightarrow -\lambda$  and $1\leftrightarrow 2$. In order to combine the terms conveniently, we write this term as
\begin{eqnarray}\label{integraltwo}
\mathcal{I}_2 & = & 16\pi y_1y_2\int\frac{d^2k}{(2\pi)^2}e^{ik\cdot z_{12}}\int_0^\infty du dt\cos(|\vec{k}|y_1\sinh u)\sin(|\vec{k}|y_2\sinh t)\cr
& \times & \frac{\frac{1}{2}-i\lambda}{\frac{1}{4}+\lambda^2}\cosh\left[\left(\frac{1}{2}+i\lambda\right)u\right]\sinh\left[\left(\frac{1}{2}+i\lambda\right)t\right].
\end{eqnarray}
The completeness relation involves a weighted integral over $\lambda$ of the sum of these two terms
\begin{eqnarray}
\mathcal{I} & = & \int_{-\infty}^\infty d\lambda \mu(\lambda) [\mathcal{I}_1+\mathcal{I}_2]\cr
& = & 16\pi y_1y_2\int\frac{d^2k}{(2\pi)^2}e^{ik\cdot z_{12}}\int_0^\infty du dt\cos(|\vec{k}|y_1\sinh u)\sin(|\vec{k}|y_2\sinh t)\mathcal{I}_\lambda.
\end{eqnarray}
We deduce that the integration weight is 
\begin{equation}
\mu(\lambda) = \frac{\mu_0\Gamma(\frac{3}{2}+i\lambda)\Gamma(\frac{3}{2}-i\lambda)}{\Gamma(\frac{1}{2}+i\lambda)\Gamma(\frac{1}{2}-i\lambda)}.
\end{equation}
The $\lambda$ integral, which is over the terms in the second lines of~\eqref{integralone} and~\eqref{integraltwo}, can be done using trigonometric sum and difference identities and using that $\int_{-\infty}^\infty d\lambda \cos(\lambda x) = 2\pi\delta(x)$. We find that it is
\begin{equation}
\mathcal{I}_\lambda = -2\pi\mu_0\partial_t\delta(u-t).
\end{equation}
Now we can substitute this into the integral and perform an integration by parts in $t$ to obtain
\begin{eqnarray}
\mathcal{I} & = & 32\mu_0y_1y_2^2\pi^2\int\frac{d^2k}{(2\pi)^2}|\vec{k}|e^{ik\cdot z_{12}}\int_0^\infty du\cos(|\vec{k}|y_1\sinh u)\cos(|\vec{k}|y_2\sinh u)\cosh u\cr
& = & 8\mu_0y_1y_2^2\pi^2\int \frac{d^2k}{(2\pi)^2} e^{ik\cdot z_{12}}\int_{-\infty}^\infty dv[\cos(v(y_1-y_2))+\cos(v(y_1+y_2))]\cr
& = & 16\mu_0y_1y_2^2\pi^3\int\frac{d^2k}{(2\pi)^2}e^{ik\cdot z_{12}}[\delta(y_1-y_2)+\delta(y_1+y_2)]\cr
& = & 16\mu_0 y_1^3\pi^3\delta(y_1-y_2)\delta^{(2)}(z_1-z_2).
\end{eqnarray}
where we have let $v=\sinh u$ and only kept $\delta(y_1-y_2)$ since $y_1,y_2>0$. The Lorentz invariant delta function in hyperbolic coordinates is 
\begin{equation}
\delta^{(3)}(\hat{p}_1,\hat{p}_2) = (2\pi)^32\hat{p}^0\delta^{(3)}(\hat{p}_1-\hat{p}_2) = 4y_1^3(2\pi)^3\delta(y_1-y_2)\delta^{(2)}(z_1-z_2)
\end{equation}
so if $\mu_0=2$ then
\begin{equation}
\mathcal{I} = \delta^{(3)}(\hat{p}_1,\hat{p}_2).
\end{equation}
\bibliography{CPBfermions}
\bibliographystyle{JHEP}
\end{document}